%% file: root.tex
\documentclass{ifacconf}

\usepackage{graphicx}      
\usepackage{natbib}
\providecommand{\main}{.}
\usepackage{\main/style/style}

\begin{document}
\begin{frontmatter}

\title{Distributed Gaussian Process Based Cooperative Visual Pursuit Control for Drone Networks} 

\thanks[footnoteinfo]{© 2022 the authors. This work has been accepted to IFAC for publication under a Creative Commons Licence CC-BY-NC-ND.}

\author[First]{Makoto Saito} 
\author[First]{Junya Yamauchi} 
\author[First]{Tesshu Fujinami} 
\author[First]{Marco Omainska} 
\author[First]{Masayuki Fujita}

\address[First]{The University of Tokyo, Tokyo 113-8656, Japan (e-mail: junya\_yamauchi@ ipc.i.u-tokyo.ac.jp).}

\subfile{sections/abstract}

\begin{keyword}
Vision-based control, cooperative control, machine learning, Gaussian process.
\end{keyword}

\end{frontmatter}

\subfile{sections/section_introduction}
\subfile{sections/section_problem_setting}
\subfile{sections/section_dGP}
\subfile{sections/section_simulation}
\subfile{sections/section_experiment}
\subfile{sections/section_conclusion}

\bibliographystyle{ifacconf}
\bibliography{root}             

\end{document}

%% file: sections/abstract.tex
\begin{abstract}
In this paper, we propose a control law for camera-equipped drone networks to pursue a target rigid body with unknown motion
based on distributed Gaussian process.
First, we consider the situation where each drone has its own dataset, and learned the unknown target motion in a distributed manner.
Second, we propose a control law using the distributed Gaussian processes, and show that the estimation and control errors are ultimately bounded.
Furthermore, the effectiveness of the proposed method is verified by using simulations and experiments with actual drones.
\end{abstract}

%% file: sections/section_introduction.tex
\section{Introduction}

With the improvement of image processing technologies such as machine learning, vision sensors are becoming increasingly important for the realization of autonomous mobile robots.
Applications of autonomous mobile robots based on visual information include teaming with humans \citep{Aggravi:2021}, bird control at airports \citep{Paranjape:2017tro}, and searching at disaster sites \citep{Chung:2011gn}.
Furthermore, networking multiple autonomous mobile robots is expected to lead to more advanced autonomous systems \citep{Chung:2018tro}.

This paper addresses cooperative control of mobile robots based on visual information.
Formation control is a typical example of a problem handled by cooperative control based on visual information.
\citep{Montijano:2016tro,Vela:2017jm}.
Although target object tracking based on visual information is a task that has many points in common with formation control, 
it is essentially different in that it involves the uncertainty of object motion.
In the target tracking problem, unknown object motion has conventionally been considered as a disturbance and tackled as a disturbance regulation problem \citep{hatanaka2015passivity}.
With the recent development of machine learning, methods for learning target motions from data using Neural Networks \citep{Harris:2020cdc} and Gaussian process regression \citep{yamauchi2021cooperative} have been reported.

In addition to being a highly expressive learning method, Gaussian process regression is a Bayesian method, which means that it can estimate learning uncertainty. 
On the other hand, Gaussian process regression is known to have high computational complexity, which increases the computational load when training with a large amount of data \citep{shen2005fast}.
To address this problem, a distributed Gaussian process \citep{deisenroth2015distributed} has been developed to reduce the computational load by splitting the dataset and learning.
Our previous work \citep{yamauchi2021cooperative} assumed that all robots had the same dataset. Therefore, it was necessary to consider the trade-off between prediction performance and computational load when the target object moves over a wide range.


In this paper, we introduce a distributed Gaussian process \citep{deisenroth2015distributed} to the results of previous work \citep{yamauchi2021cooperative} and propose a cooperative visual tracking control law based on distributed learning.
Here, we consider a situation in which each robot equipped with a visual sensor learns a target motion model based on a different dataset using a Gaussian process.
Then, a control law is derived using the prediction of the target motion integrated within the network according to the uncertainties in learning.
Next, we show that the pursuit performance of each robot is ultimately bounded with high probability by the proposed control law.
Finally, the effectiveness of the proposed control law is verified by simulation and experiment.
The contributions of this paper are threefold: (i) the dataset can be divided and held by each drone to reduce the computational load during learning and target motion prediction, (ii) ultimate boundedness of estimation and control error is theoretically shown even when the dataset is divided, and (iii) the effectiveness of the proposed method is verified by experiments. 


%% file: sections/section_problem_setting.tex
\section{problem setting}
\subsection{Rigid Body Motion and Network}
\begin{figure}
    \begin{center}
        \includegraphics[width=0.75\linewidth]{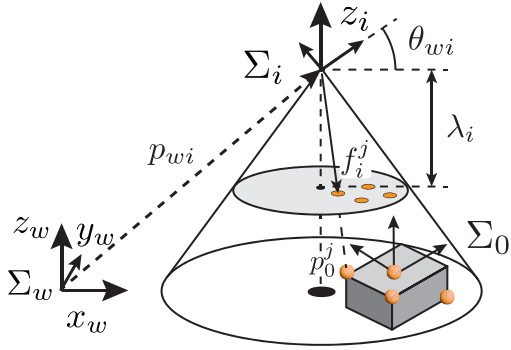}
    \end{center}
    \caption{Coordinate of drones and visual measurements.}
    \label{fig:coordinate}
\end{figure}

In this paper, we consider the case where the target and $n$ drones $\mathcal{V}\coloneqq\{1, \dots, n\}$ are all rigid bodies.
The target rigid body is labeled as $0$.
Here, assuming that the coordinate frame $\Sigma_i$ is attached to the rigid body. 
The $i$-th rigid body motion is described using the relative position $p_{wi} \in \R^3$ and relative orientation $R(\theta_{wi}) \in \R^{3 \times 3}$ of $\Sigma_i$ from the world coordinate frame $\Sigma_w$,
where $\theta_{wi} \in (-\pi, \pi]$ is the rotation angle around $z_w$ axis as shown in Fig. \ref{fig:coordinate}.

The pose of the $i$-th rigid body is denoted by $g_{wi}= (p_{wi}, R(\theta_{wi}))$ seen from the world coordinate $\Sigma_w$.
We also describe the pose $g_{wi}$ by the following matrix form:
\begin{align}\label{equ:gwi}
    g_{wi} &= \left[
        \begin{array}{cc}
            R(\theta_{wi}) & p_{wi} \\
            0      & 1
        \end{array}
        \right], \quad i\in \{0\} \cup \mathcal{V}, \\
    R(\theta_{wi}) &=
    \left[\begin{array}{ccc}
        \cos(\theta_{wi}) & -\sin(\theta_{wi}) & 0 \\
        \sin(\theta_{wi}) & \cos(\theta_{wi}) & 0 \\
        0 & 0 & 1
    \end{array}\right].
\end{align}

Let us define $\Vb{wi} \coloneqq \left[(v^b_{wi})^T~ \omega^b_{wi}\right]^T \in \R^4$ as the body velocity of the $i$-th rigid body, 
where $v^b_{wi} \in \R^3$ is the linear body velocity and 
$\omega^b_{wi} \in \R$ is the angular body velocity.
The rigid body motion can then be described as
\begin{equation}
    \dot{g}_{wi} = g_{wi}\hVb{wi}, \quad \hVb{wi} \coloneqq
    \left[
        \begin{array}{cc}
            \hat{\omega}_{wi}^b & v^b_{wi} \\
            0                   & 0
        \end{array}
        \right] \in \mathbb{R}^{4 \times 4}, \label{equ:rigidbody_motion}
\end{equation}
where the operator $\wedge$ is defined by 
\begin{align}
    \hat{a} = \left[
    \begin{array}{ccc}
        0 & -a & 0 \\ a & 0 & 0 \\ 0 & 0 & 0 
    \end{array}
    \right], \quad a \in \R.
\end{align}

Using (\ref{equ:rigidbody_motion}), we can also describe the motion of the relative pose $g_{ij} \coloneqq g^{-1}_{wi} g_{wj}$ of rigid body $i$ and rigid body $j$.

In this paper, we assume that the pose of the target $g_{w0}$ cannot acquired by any drone although each drone can obtain the information of its own pose $g_{wi}$ by an external localization system. In addition, we impose the assumption that the target is moving in a bounded region.
\subfile{../assumptions/ass_target_motion}

The group of $n$ drones can communicate with neighboring drones.
This network is represented by $\mathcal{G}=(\mathcal{V}, \mathcal{E}),~ \mathcal{E} \subseteq \mathcal{V} \times \mathcal{V}$.
For this network, we assume the following:
\subfile{../assumptions/ass_graph}
The graph Laplacian matrix associated with the graph $\mathcal{G}$ is denoted as $L$. 
For more details of the graph theory, refer to \citep{mesbahi2010graph}.

\subsection{Online Data: Visual Measurements}
Suppose that the target rigid body has $m$ feature points and their positions are expressed as $p_0^i \in \R^3, i \in \{1, \dots, m\}$ in the target coordinate system $\Sigma_0$.
Denoting the position of the $j$-th feature point of the target viewed from the camera coordinates of the $i$-th drone as $p_i^j \coloneqq [x_i^j~ y_i^j~ z_i^j]^T$, the coordinates of the $j$-th feature point on the camera of $i$-th drone denoted as
\begin{equation}\label{equ:f}
    f_i^j = \frac{\lambda_i}{z_i^j} \left[
        \begin{array}{c}
            x_i^j \\
            y_i^j
        \end{array}
        \right],
\end{equation}
where $\lambda_i$ is the focal length of the camera of the $i$th drone.
We also define $f_i \coloneqq [(f_i^1)^T ~ \cdots ~ (f_i^m)^T ]^T$.
In this paper, let $\mathcal{V}_v$ denote the set of drones that observe all feature points, and we call it \textit{visibility set}.
We make the following assumptions on the visibility set $\mathcal{V}_v$.
\subfile{../assumptions/ass_v}
This assumption means that one or more drones can observe all feature points of the target.

Furthermore, using
\begin{equation}
    v_i \coloneqq
    \left\{
    \begin{array}{ll}
        1 & \textrm{if} \quad i \in \mathcal{V}_v \\
        0 & \textrm{otherwise}
    \end{array}
    \right. ,
\end{equation}
and the graph Laplacian $L$, we define $H$ as follows:
\begin{align}
    D_v &= \diag{d_1v_1, \dots , d_nv_n} \in \mathbb{R}^{n \times n}, \\
    H &\coloneqq L + D_v,
\end{align}
where $d_i > 0$.
In this case, by \citep{ni2010leader}, the following lemma is derived.
\subfile{../lemmas/lemma_weighted_H}

\subsection{Offline Training Data}
In this paper, we assume that each drone has a different dataset for learning the target body velocity $\Vb{w0}$.
Assume that the body velocity of the target can be observed as
\begin{equation}\label{equ:observe_vel}
    y = \Vb{w0}(\check{g}_{w0}) + \epsilon, \quad 
    \check{g}_{wi} \coloneqq [p_{wi}^T, \theta_{wi}]^T, 
\end{equation}
where $\epsilon \coloneqq [\epsilon_1~ \dots~ \epsilon_4]^T \in \mathbb{R}^4$, $\epsilon_i \sim \mathcal{N}(0, \sigma_{ni}^2)$ is sub-Gaussian noise. 

The $i$-th drone has $M_i$ data pair $(y, \check{g}_{wi})$ in advance.
Then, a dataset for each drone is defined as
\begin{align}
    \begin{split} \label{equ:gp_data}
            X^{(i)} & \coloneqq [\check{g}_{w0}^{(i1)}, \dots ,\check{g}_{w0}^{(iM_i)}] \in \R^{4 \times M_i}, \\
        Y^{(i)} & \coloneqq [y^{(i1)}, \dots , y^{(iM_i)}] \in \R^{4\times M_i}, \\
        D^{(i)} & \coloneqq \{X^{(i)}, Y^{(i)}\}.
    \end{split}
\end{align}

\subsection{Control Objective }
At the end of this section, we formulate the problem to be considered in this paper.
First, we introduce desired relative poses $g_{di} = (p_{di}, R(\theta_{di}))$ for each drone, which is constant ($\dot{g}_{di}\equiv0$).
Then, we define the control error $\widetilde{e}_{ci}$ as (\ref{equ:g_ci_desired}).
\begin{equation} \label{equ:g_ci_desired}
    \widetilde{g}_{ci} \coloneqq g_{di}^{-1} g_{i0}, \quad
    \widetilde{e}_{ci}\coloneqq vec(\widetilde{g}_{ci}),
\end{equation}
where $vec$ is defined as follows:
\begin{align}
    vec(g) &\coloneqq [p^T~ \sin \theta]^T \in \R^4.
\end{align}

However, since $g_{i0}$ is unknown, we redefine the control error $e_{ci}$ as in (\ref{equ:g_ci}) using $\bar{g}_{i0}$, the estimated value of $g_{i0}$.
\begin{equation}\label{equ:g_ci}
    g_{ci} \coloneqq g_{di}^{-1} \bar{g}_{i0}, 
    \quad e_{ci} \coloneqq vec(g_{ci}).
\end{equation}

In order to estimate $\bar{g}_{i0}$, we define the estimation error $e_{ei}$ as
\begin{equation}\label{equ:g_ei}
    g_{ei} \coloneqq \bar{g}_{i0}^{-1} g_{i0}, 
    \quad e_{ei} \coloneqq vec(g_{ei}).
\end{equation}

Although $g_{i0}$ is unknown, for $m \ge 4$, $e_{ei}$ can be obtained as $e_{ei} = J_i^\dagger (f_i - \bar{f}_i)$ with the pseudo-inverse matrix of image Jacobian $J_i^\dagger$ \citep{hatanaka2015passivity}.

In this paper, we make the following assumption on the estimation error $\theta_{ei}$ and control error $\theta_{ci}$.
\subfile{../assumptions/ass_est_bound}

From the above, the system can be described by (\ref{equ:ei}),(\ref{equ:ci}).
\begin{subequations} \label{eq:e}
    \begin{align}
        \dot{g}_{ei} &= \hat{u}_{ei}g_{ei} + g_{ei}\hVb{w0} \label{equ:ei},\\
        \dot{g}_{ci} &= \hat{u}_{ci}g_{ci} - g_{ci}\hat{u}_{ei} \label{equ:ci},
    \end{align}
\end{subequations}
where $u_{ei}$ and $u_{ci}$ are an observer and  control input, respectively.

In addition, the error of the entire network $e$ is determined as follows:
\begin{align}
    e_c&\coloneqq[e_{c1}^T~ \dots e_{cn}^T]^T, \quad 
    e_e\coloneqq[e_{e1}^T~ \dots~ e_{en}^T]^T, \\
    e_i&\coloneqq[e_{ci}^T~ e_{ei}^T]^T \in \mathbb{R}^{8}, \quad
    e \coloneqq [e_c^T~ e_e^T]^T \in \mathbb{R}^{8n}.
\end{align}

The goal of this paper is to propose a control law that achieves the following:
\begin{align} \label{equ:goal}
    \lim_{t\rightarrow \infty}\|e(t)\| \le b,\quad b \ge 0.
\end{align}

%% file: assumptions/ass_target_motion.tex
\begin{assumption}\label{ass:target_motion}
    The trajectory of the target position 
    $p_{w0}(t),~ \\ \forall t \ge 0$ belongs to a compact set $\mathcal{X} \subset \mathbb{R}^3$. 
\end{assumption}

%% file: assumptions/ass_graph.tex
\begin{assumption}\label{ass:graph}
    The graph $\mathcal{G}$ is undirected and fixed.
\end{assumption}

%% file: assumptions/ass_v.tex
\begin{assumption}\label{ass:v}
    The visibility set $\mathcal{V}_v$ is not an empty set.
\end{assumption}

%% file: lemmas/lemma_weighted_H.tex
\begin{lem}\label{lem:weighted_H}
    When Assumption \ref{ass:v} holds, the matrix $H$ is a positive definite matrix.
\end{lem}

%% file: assumptions/ass_est_bound.tex
\begin{assumption}\label{ass:est_bound}
     The control error angles $\theta_{ci}(t)$ and estimation error angles $\theta_{ei}(t)$ for all $i \in \V$ are bounded such that $\theta_{ci}(t)\in (-\pi/2, \pi/2)$, $\theta_{ei}(t)\in (-\pi/2, \pi/2)$ for all $t \geq 0$.
\end{assumption}

%% file: sections/section_dGP.tex
\section{Visual Pursuit Control with Distributed Gaussian Process}
\subsection{Distributed GP}
Since we consider pursuing a moving target, it is necessary to predict the body velocity of the target.
In this paper, we employ a distributed Gaussian process regression to estimate the body velocity.
This is to reduce the computational load on the drone, and thus 
consider the case where the $i$-th drone possesses a GP expert model that performs Gaussian process regression using $D^{(i)}$
\citep{deisenroth2015distributed}.

As input to the Gaussian process regression, we use $\bar{g}_{w,i} \coloneqq g_{wi}\bar{g}_{i0}$.
This is the estimate of $g_{w0}$ by $i$-th drone.
We denote the predicted distribution of the $i$-th element of $\Vb{w0}$ by the $j$-th drone as $\mathcal{N}(\mu_i^{(j)}, var_i^{(j)})$.
Then, the predicted body velocity at a new pose $\check{g}_{w0}^*$ using Gaussian process model is obtained as follows:
\begin{align}
    \mu_i^{(j)}
    &= (k_{i*}^{(j)})^T(\mathcal{K}_i^{(j)}+\sigma_{ni}^2 I_{M_j})^{-1}Y_i^{(j)}, \\
    var_i^{(j)} 
    &= k_i^{(j)}(\check{g}_{w0}^*, \check{g}_{w0}^*) - k_{i*}^{(j)T}
    (\mathcal{K}_i^{(j)}+\sigma_{ni}^2 I_{M_j})^{-1}k_{i*}^{(j)},
\end{align}
where, using covariance function $k_i^{(j)} : \mathbb{R}^4 \times \mathbb{R}^4 \rightarrow \mathbb{R}$, $\mathcal{K}_i^{(j)}$ and $k^{(j)}_{i*}$ is defined as 
\begin{align}
    [\mathcal{K}_i^{(j)}]_{ll'} &\coloneqq k_i^{(j)}(\check{g}_{w0}^{(jl)}, \check{g}_{w0}^{(jl')}), \\
    [k_{i*}^{(j)}]_l &\coloneqq k_i(\check{g}_{w0}^*, \check{g}_{w0}^{(jl)}),
\end{align}
and, $Y_i^{(j)} \in \mathbb{R}^{M_j}$ is a vector which represents $i$-th column of $Y^{(j)}$.

In this paper, we use the SE-ARD kernel, which is a universal kernel, as the kernel function $k_j^{(i)}$.
The SE-ARD kernel is defined as below.
\begin{equation}
    k_i^{(j)}(x, x') = (\sigma_{fi}^{(j)})^2 \exp\left(
    - \sum_{j=1}^4 \frac{(x_j - x'_j)^2}{2(l_{ij}^{(j)})^2}
    \right)
\end{equation}
The hyperparameters of the kernel functions $k_i^{(j)}$ are optimized by the dataset (\ref{equ:gp_data}).

The predicted distribution of $\Vb{w0}$ is then obtained as,
\begin{align}
    \mu^{(i)}(\check{\bar{g}}_{w,i})    & \coloneqq [\mu_1^{(i)}~ \dots~ \mu_4^{(i)}]^T \in \mathbb{R}^4,               \\
    \Sigma^{(i)}(\check{\bar{g}}_{w,i}) & \coloneqq \diag{var_1^{(i)}, \dots, var_4^{(i)}} \in \mathbb{R}^{4 \times 4}.
\end{align}

In this paper, we consider combining the results of Gaussian process regression with neighboring drones in the network by using the product-of-GP-experts model with reference to \citep{deisenroth2015distributed}.
In this case, we obtain the following as mean and variance of the predictive distribution of the $j$-th element of $\Vb{w0}$ by the $i$-th drone.
\begin{align}
    \mu_j^{poe(i)}        &\coloneqq
    \sum_{k \in \mathcal{N}_i \cup \{i\}}
    w_{ijk} \mu_j^{(k)} (\check{\bar{g}}_{w,k}), \label{equ:poe_mu}                                        \\
    w_{ijk}               &\coloneqq var_j^{poe(i)} (var_j^{(k)}(\check{\bar{g}}_{w,k}))^{-1}, \\
    (var_j^{poe(i)})^{-1} &\coloneqq
    \sum_{k \in \mathcal{N}_i \cup \{i\}}
    (var_j^{(k)} (\check{\bar{g}}_{w,k}))^{-1},  \label{equ:poe_var}
\end{align}
where $\mathcal{N}_i$ is the set of drones adjacent to the $i$-th drone.

From the discussion above, we treat the following as the mean and variance of the predictive distribution of $\Vb{w0}$ for the $i$-th drone.
\begin{align}
    \mu^{poe(i)}    & \coloneqq [\mu_1^{poe(i)}, \dots, \mu_4^{poe(i)}]^T \in \mathbb{R}^4,              \\
    \Sigma^{poe(i)} & \coloneqq \diag{var_1^{poe(i)}, \dots, var_4^{poe(i)}} \in \mathbb{R}^{4 \times 4}.
\end{align}

The computational cost of learning a Gaussian process model is $O((\sum_{i \in \mathcal{V}} M_i)^3)$ when using a single Gaussian process regression, but the overall computational cost of learning a Gaussian process model can be reduced to $O(\sum_{i \in \mathcal{V}} M_i^3)$ using a distributed Gaussian process regression. 
Although the computational effort for prediction is the same, both learning and prediction are performed in parallel for each GP expert, which allows the computation to be completed in a shorter time and with less load on each computer. 
This is useful for drones, for which it is difficult to install high-performance computers due to power consumption and weight. In addition, it is expected to improve tracking performance by reducing the delay due to computation time, even in the case of actual control. 

\subsection{Bound on Prediction Error}
In this paper, we predict the target body velocity using a Gaussian process.
At this point, it is necessary to consider the upper bound of the prediction error when discussing the stability of the control.


For Gaussian process regression, the following inequality holds for the prediction error \citep[theorem 6]{srinivas2012information}.
\subfile{../lemmas/lemma_gp_error_bound}
With this lemma, we were able to determine the upper bound when single Gaussian process regression is used.

However, in this paper, since the Gaussian process regression results are combined to obtain the distributed Gaussian process regression results, it is necessary to obtain an upper bound on this as well.
To calculate the upper bound,we define $L_{\mu_j}^{(i)}$ as the Lipschitz constant of $\mu^{(j)}_i$ from \citep{rasmussen2003gaussian}, and also define
\begin{align}
    L_{\mu_j} \coloneqq \max_{i \in \mathcal{V}} L_{\mu_j^{(i)}}, \quad
    \bar{\Delta}_j \coloneqq \max_{i \in \mathcal{V}} \bar{\Delta}_j^{(i)}.
\end{align}
Furthermore, we also define $L_\mu$ and $\bar{\Delta}$ as
\begin{align}
    L_\mu \coloneqq \|[L_{\mu_1} \cdots L_{\mu_4}]^T\|, \quad 
    \bar{\Delta} \coloneqq \|[\bar{\Delta}_1 \cdots  \bar{\Delta}_4]^T\|.
\end{align}
Then from the Lemma \ref{lem:gp_error_bound}, about the prediction error of single Gaussian process regression, the following lemma holds.
\subfile{../lemmas/lemma_dgp_error_bound}

This lemma allows us to obtain an upper bound on the prediction error as in Lemma \ref{lem:gp_error_bound}, even when using a distributed Gaussian process regression.
Using this upper bound, we will discuss the stability of the control law using the distributed Gaussian process proposed hereafter.

\subsection{Control Law}
A passivity-based visual pursuit control law has been proposed in previous studies.
Based on the passivity, this paper also proposes a control law using distributed Gaussian process regression.

First, let us define the storage function $S_i$ for the $i$-th drone as follows:
\begin{equation}
    S_i \coloneqq \frac{1}{2}\|p_{ei}\|^2
    + \frac{1}{2}\phi(R(\theta_{ei}))
    + \frac{1}{2}\|p_{ci}\|^2
    + \frac{1}{2}\phi(R(\theta_{ci})).
\end{equation}
where $\phi(R) \coloneqq (1/2)\|I_3 - R\|^2_F$ and $\|\cdot\|_F$ is the Frobenius norm.

Furthermore, the following lemma holds for the time derivative of $S_i$ from \citep{hatanaka2015passivity}
\subfile{../lemmas/lemma_S}
where $\Ad{R(\theta)}$ is called adjoint transformation and defined as
\begin{align}
    \Ad{R(\theta)} &\coloneqq \diag{R(\theta), 1}. 
\end{align}

Here, we propose the following control law which estimates the target motion by distributed Gaussian process as in the case of \citep{yamauchi2021cooperative} from passivity.

\begin{equation}
    u_i = -K_i N_i e_i
    - B_i \sum_{j \in N_i} k_s
    vec(\bar{g}_{w,i}^{-1}\bar{g}_{w,j})
    - A_i\mu^{poe(i)}. \label{equ:ui}
\end{equation}
Here, each matrix is defined as follows:
\begin{align}
    N_i & \coloneqq
    \left[
        \begin{array}{cc}
            I_4                     & 0   \\
            -\Ad{R(\theta_{ci})} & I_4
        \end{array}
    \right] \in \mathbb{R}^{8 \times 8},                                         \\
    B_i & \coloneqq
    \left[
        \begin{array}{cc}
            \Ad{R^T(\theta_{ci})} & I_4
        \end{array}
    \right]^T \in \mathbb{R}^{8 \times 4},                                        \\
    A_i & \coloneqq
    \left[
    \Ad{R^T(\theta_{ei})} \Ad{R^T(\theta_{ci})} \quad
    \Ad{R^T(\theta_{ei})}
    \right] ^T \in \mathbb{R}^{8 \times 4},                                       \\
    K_i & \coloneqq \diag{k_{ci1}, \dots, k_{ci4}, v_i k_{ei1}, \dots, v_ik_{ei4}} \in \mathbb{R}^8.
\end{align}

The proposed control law replaces the Gaussian process regression in the conventional control law \citep{yamauchi2021cooperative} with a distributed Gaussian process regression.
Conventionally, the variance of the prediction distribution, which can be used to evaluate the reliability of Gaussian process regression, is used to change the magnitude of the gain, but in this control law, it is used to determine the weight when combining the prediction results of neighboring drones.

\begin{figure}
    \begin{center}
        \includegraphics[width=\linewidth]{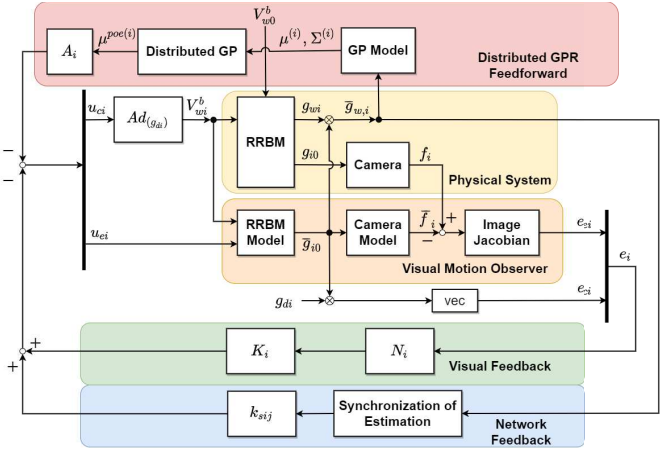}
    \end{center}
    \caption{Block diagram of the proposed control law}
    \label{fig:block}
\end{figure}

A block diagram of this control system is shown in Fig.\ref{fig:block}.

\subsection{Stability Analysis}
In this section, we analyze the stability of the control law proposed in (\ref{equ:ui}).
First, the storage function of the entire system is defined as follows:
\begin{align} \label{eq:S}
    S \coloneqq \sum_{i \in \V}S_i.
\end{align}
Then, by \citep{yamauchi2021cooperative}, we can prove the following lemma.

\subfile{../lemmas/lemma_network_S}
\begin{proof}
    See the proof of \cite[Theorem 2]{yamauchi2021cooperative}.
    \qed
\end{proof}

Using the Lemma \ref{lem:network_S}, we obtain the main theorem.
Before we show the theorem, we define the following:
\begin{align}
    P^i  = \left[ P^i \right]_{jk} \coloneqq \left\{
    \begin{array}{ll}
        1   & i = j = k                           \\
        0.5 & i = j ~ \textrm{or} ~ i = k \\
        0   & \textrm{otherwise}
    \end{array}
    \right. \in \mathbb{R}^{n\times n} \label{equ:p_i}.
\end{align}

\subfile{../theorems/theorem_distributedGP}

Theorem \ref{thm:dGP} implies that (\ref{equ:goal}) is probabilistically satisfied by (\ref{equ:dgp_B}).

Therefore, we have newly shown that it is possible to reduce the computational load of Gaussian process regression and the communication load for sharing the observed training data with other computers, and that it is possible to perform stable control as in the case of using a single Gaussian process model.

%% file: lemmas/lemma_gp_error_bound.tex
\begin{lem}\label{lem:gp_error_bound}
    Consider the error system (\ref{eq:e}) and Gaussian process models trained by the dataset in (\ref{equ:gp_data}). 
    Then, the model error satisfies the following:
    \begin{equation}
        P\{\forall x \in \mathcal{X} \times (-\pi, \pi] \mid \|(V^b_{w0}(x))_i - \mu^{(j)}_i(x)\|
        \le  \bar{\Delta}_i^{(j)} \}
        \ge \delta^{1/4},
    \end{equation}
    where, $\delta \in (0, 1)$
    $\bar{\Delta}_i^{(j)} := \beta_i^{(j)} (var_i^{(j)})^{\frac{1}{2}}$, \\
    $\beta_i^{(j)} =
        \sqrt{
            2 \|(V^b_{w0})_i\|_{k_i^{(j)}}^2 + 300 \gamma_i^{(j)} \ln^3 \left(\frac{M^{(j)} + 1}{1 - \delta^{1/4}}\right)
    }$ and $\gamma_i^{(j)}$ is maximum information gain.
\end{lem}

%% file: lemmas/lemma_dgp_error_bound.tex
\begin{lem}\label{lem:dgp_error_bound}
    Assume that each drone has a dataset (\ref{equ:gp_data}) and is able to compute (\ref{equ:poe_mu}), by communicating with neighboring drones by the graph $\G$ satisfying Assumption \ref{ass:graph}.
    Then, the following inequality holds with probability $\delta \in (0, 1)$ with
    $\|e_{e}^{max}\| := \max_{i \in \mathcal{V}} \|e_{ei}\|$.
    \begin{align}
        \|V^b_{w0}(\check{g}_{w0}) - \mu_i^{poe(i)}(\check{\bar{g}}_{w,i})\| \le \bar{\Delta} + \frac{\pi}{2} L_\mu\|e_{e}^{max}\|
    \end{align}
\end{lem}
\begin{proof}
    Define $W_{ij} \coloneqq \diag{w_{i1j}, \dots , w_{i4j}}$.
    Using the triangle inequality, we derive the following:
    \begin{align}
         & \| V^b_{w0}(\check{g}_{w0}) - \mu^{poe(i)} \|
        \\
         & \le \|
        \sum_{j \in \mathcal{N}_i \cup \{i\}}
        W_{ij} (V^b_{w0}(\check{g}_{w0}) -\mu^{(j)}(\check{g}_{w0})) \|
        \\
         & + \|
        \sum_{j \in \mathcal{N}_i \cup \{i\}}
        W_{ij} (\mu^{(j)}(\check{g}_{w0}) - \mu^{(j)}(\check{\bar{g}}_{w,j}))
        \|.
    \end{align}
    Since
    $
        \sum_{j \in \mathcal{N}_i \cup \{i\}}
        W_{ij} = I_4
    $
    and with \cite[Lemma 4.2]{omainska2021gaussian},
    \cite[Proposition C.3(ii)]{hatanaka2015passivity}, we have
    \begin{align}
        &\| \sum_{j \in \mathcal{N}_i \cup \{i\}}
        W_{ij} (V^b_{w0}(\check{g}_{w0}) -\mu^{(j)}(\check{g}_{w0})) \| 
        \le \bar{\Delta} \\
        &\|
        \sum_{j \in \mathcal{N}_i \cup \{i\}}
        W_{ij}(\mu^{(j)}(\check{g}_{w0}) - \mu^{(j)}(\check{\bar{g}}_{w,j}))
        \| \\
        &\hspace{5mm}\le L_\mu \|[(p_e^{max})^T, \theta_e^{max}]^T\| \le \frac{\pi}{2} L_\mu \|e_e^{max}\|.
    \end{align}    
    From the above, we can derive
    \begin{equation}
        \| V^b_{w0}(\check{g}_{w0}) - \mu^{poe(i)}(\check{\bar{g}}_{w,i})) \|
        \le 
        \bar{\Delta} + \frac{\pi}{2} L_\mu \|e_e^{max}\|.
    \end{equation}
    This completes the proof.
    \qed
\end{proof}

%% file: lemmas/lemma_S.tex
\begin{lem}\label{lem:S}
    The time derivative of the storage function $S_i$ can be calculated as follows:
    \begin{equation} \label{equ:dot_S}
        \dot{S}_i = e_i^T N_i^T u_i + e_i^T
        \left[
            \begin{array}{c}
                0 \\
                \Ad{R(\theta_{ei})}
            \end{array}
            \right]
        V_{w0}^b,
    \end{equation}
\end{lem}

%% file: lemmas/lemma_network_S.tex
\begin{lem}\label{lem:network_S}
    When the target is static i.e. $\Vb{w0} \equiv 0$, the time derivative of the storage function $S$ in (\ref{eq:S}) obeys the following:
    \begin{equation}
        \dot{S} \le -P_e^T H P_e - \Phi_e^T H \Phi_e - e_c^T \Lambda e_c,
    \end{equation}
    where $\lambda_{min}(A)$ is the minimum eigenvalue of $A$, each notation above is defined as follows:
    \begin{align}
        \lambda_{Ki} &=
        \left\{
        \begin{array}{ll}
            \lambda_{min}(N_i^T K_i N_i) & \textrm{if} \quad i \in \V_v \\
            \lambda_{min}(K_c)           & \textrm{otherwise}
        \end{array}
        \right. ,\\
        D_v &:= \diag{v_1 \lambda_{K1}, \dots, v_n \lambda_{Kn}}, \\
        \Lambda &:= \diag{\lambda_{K1},\dots,\lambda_{Kn}} \otimes I_4, \\
        P_e &:= [\|p_{e1}\| \dots  \|p_{en}\|]^T, \\
        \Phi_e &:= \left[\sqrt{\phi(R(\theta_{e1}))}
                \dots \sqrt{\phi(R(\theta_{en}))} \right]^T.
    \end{align}
\end{lem}

%% file: theorems/theorem_distributedGP.tex
\begin{thm}\label{thm:dGP}
    Consider the system (\ref{equ:ei}), (\ref{equ:ci}), (\ref{equ:ui}) with a learned Gaussian process model obtained from the dataset in (\ref{equ:gp_data}).
    Suppose that Assumptions \ref{ass:target_motion}--\ref{ass:est_bound} hold.
    Let $\lambda_{H}$ be the minimum eigenvalue of $H$ and $\lambda_{P}$ be the maximum eigenvalue of $P^i$.
    Suppose that $\lambda_{H}$ and $\lambda_{P}$ satisfies 
    \begin{equation}\label{equ:gain_condition}
        \kappa:=\lambda_{H} - \frac{\pi}{2} L_\mu \lambda_{P}
        - \frac{1}{2\gamma^2} > 0, \quad \gamma >0.
    \end{equation}
    Then with at least probability $\delta \in (0,1)$ and $\rho(\delta) > 0,~ T(\delta) \ge 0$, 
    \begin{align}
        &P\{\|e(t)\| \le b,~ \forall t \ge T \} \ge \delta, \\
        &b(\delta) \coloneqq
        \gamma \overline{\Delta} \sqrt{\frac{n}{2\underline{\kappa}\eta}}, \quad 
        \eta \in (0,1) \label{equ:dgp_B}
    \end{align} holds for $\forall e(0)$ such that $\|e(0)\| \le \rho(\delta)$.
    where $\underline{\kappa}$ is the minimum value of $\kappa$ and the eigenvalue of $\Lambda$.
\end{thm}
\begin{proof}
    Using Lemma \ref{lem:S} and \ref{lem:network_S}, we have,
    \begin{align}
        \dot{S}
        \le& -P_e^T H P_e - \Phi_e^T H \Phi_e - e_c^T \Lambda e_c \\
        &+ \sum_{i \in \mathcal{V}} e_i^T
        \left[
        \begin{array}{c}
            0 \\
            \Ad{R(\theta_{ei})}
        \end{array}
        \right]
        (V^b_{w0}(p_{w0}) - \mu^{poe(i)})                        \\
        \le& -P_e^T H P_e - \Phi_e^T H \Phi_e - e_c^T \Lambda e_c \\
        &+ \sum_{i \in \mathcal{V}} \| e_{ei} \|\| V^b_{w0}(p_{w0}) - \mu_i^{poe} \|. \label{equ:dgp_equ_1}
    \end{align}
    Then, the following inequality holds with probability $\delta$ from Lemma \ref{lem:dgp_error_bound} and Peter-Paul's inequality.
    \begin{align}
        &\sum_{i \in \mathcal{V}} \| e_{ei} \|\| V^b_{w0}(p_{w0}) - \mu_i^{poe} \|  \\
        \le&
        \sum_{i \in \mathcal{V}} \| e_{ei} \|
        (\bar{\Delta}
        + \frac{\pi}{2} L_{\mu} \|e_e^{max}\|)
        \\
        \le&
        \sum_{i \in \mathcal{V}}
        \left(
        \frac{1}{2\gamma^2} \| e_{ei} \|^2 +
        \frac{\gamma^2}{2} \overline{\Delta}^2 + \frac{\pi}{2} L_\mu\|e_{ei}\|\|e_e^{max}\|
        \right)  \\
        =&~
        \frac{1}{2\gamma^2}\|e_e\|^2 + \frac{\pi}{2} L_{\mu}
        \max_{i\in \mathcal{V}}(v_e^T P^i v_e) + \frac{\gamma^2 n}{2} \overline{\Delta}^2.
    \end{align}
    
    Therefore, defined $v_e$ as $v_e := [\|e_{e1}\|, \dots,\|e_{en}\| ]^T$, (\ref{equ:dgp_equ_1}) can be transformed as follows:
    \begin{align}
        \dot{S}
        \le& -P_e^T H P_e - \Phi_e^T H \Phi_e - e_c^T \Lambda e_c 
        + \frac{1}{2\gamma^2}\|e_e\|^2 \\
        &+ \frac{\pi}{2} L_{\mu}\max_{i \in \mathcal{V}}
        (v_e^T P^i v_e) 
        + \frac{\gamma^2 n}{2} \overline{\Delta}^2 
        \\
        \le& - (\lambda_{H} - \frac{\pi}{2} L_\mu \lambda_{P} - \frac{1}{2\gamma^2}) \|e_e\|^2
        + \frac{\gamma^2 n}{2} \overline{\Delta}^2  
        - e_c^T \Lambda e_c.
    \end{align}
    
    From the above, we can show that
    \begin{equation} \label{equ:dgp_S_result}
        \dot{S} \le - \eta\underline{\kappa}\|e\|^2 + \frac{\gamma^2 n}{2} \overline{\Delta}^2
    \end{equation}
    when (\ref{equ:gain_condition}) holds with at least probability $\delta$.
    From (\ref{equ:dgp_S_result}), the ultimate boundedness of $e$ is shown in the same way as \cite[Theorem 3]{yamauchi2021cooperative}
    and (\ref{equ:dgp_B}) is also derived.
    \qed
\end{proof}

%% file: sections/section_simulation.tex
\section{simulation}
\begin{figure}
    \begin{center}
        \includegraphics[width=60mm]{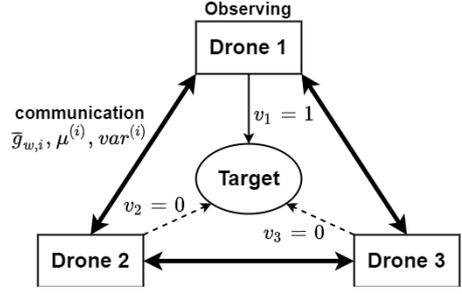}
    \end{center}
    \caption{Network of drones in simulation.}
    \label{fig:sim_system}
\end{figure}

In this simulation, we assume that the target moves according to an equation that imitates a Duffing oscillator.
The simulations are performed on three drones.
The drones is assumed to be able to communicate with two other drones, and the drone's network is shown in Fig. \ref{fig:sim_system}.

The body velocity input for the target rigid body is as follows:
\begin{align}
    V^b_{w0} &= 0.5 \Ad{R^T(\theta_{w0})} 
    \left[
        \begin{array}{c}
            y_{w0} \\
            z_{w0} \\
            v_z \\
            2\omega 
        \end{array}
    \right], \\
    v_z &\coloneqq 0.2 (-\delta z_{w0} - \alpha y_{w0} - 3 \beta x_{w0}^2 y_{w0} - \gamma \omega \sin(\theta_{w0})).
\end{align}
where $\delta = 0.1,~ \gamma = 0.39,~ \omega = 0.4,~ \alpha = -1,~ \beta = 1$ and the initial position and rotation are $p_{w0} = [-0.3~ -1.0~ 0.0]^T,~ \theta_{w0} = 0$.

Next, we describe the dataset used for training Gaussian process regression.
We use a visual motion observer \citep{hatanaka2015passivity} to calculate $\bar{g}_{w,i}$.
The dataset was created by computing an estimate of $V^b_{w0}$ by $\hat{\bar{V}}^b_{w0} = \bar{g}_{w,i}\dot{\bar{g}}_{w,i}$.
Using this estimate, we created a dataset as
\begin{equation}
    y = \bar{V}_{w0}^b(\bar{g}_{w,i}) + \epsilon,
\end{equation}
where $\epsilon \sim \mathcal{N}(0, \diag{\sigma_1^2, \dots, \sigma_4^2)},~ \sigma_i^2 = 0.01$.

The $xy$-coordinate plane is divided into three parts, each of which serves as an expert region for each drone. 
In this paper,
It is assumed that the drone can only acquire training data within the expert region.

The dataset was selected from a full 10 points within each expert region.
In the simulation, hyperparameters were learned using this data in advance.

\begin{figure}
    \begin{center}
        \includegraphics[width=75mm]{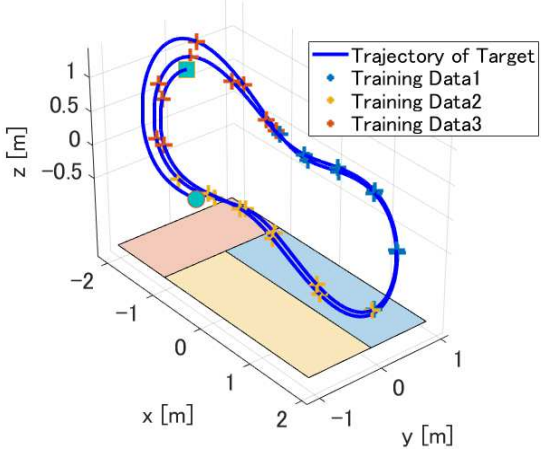}
    \end{center}
    \caption{Trajectory of the target rigid body,
    expert regions of the three drones and the training data for each drone.
    The markers `$\circ$' and `$\square$' show the starting point and  end point of the trajectory.}
    \label{fig:trajectory}
\end{figure}

In summary, the target trajectory and
the observable area and dataset for each drone are shown in Fig. $\ref{fig:trajectory}$.

The simulations were performed using a Gaussian process model trained on the dataset.
Here, the gains are set to be $k_c = k_e = 100I_6,~ k_s = 70$.
With these settings, $L_\mu=3.3, ~ \gamma^2=10$, and $\kappa=2.1$ are calculated, which satisfy the gain condition of the Theorem \ref{thm:dGP}.

\begin{figure}[htbp]
    \begin{center}
        \includegraphics[width=75mm]{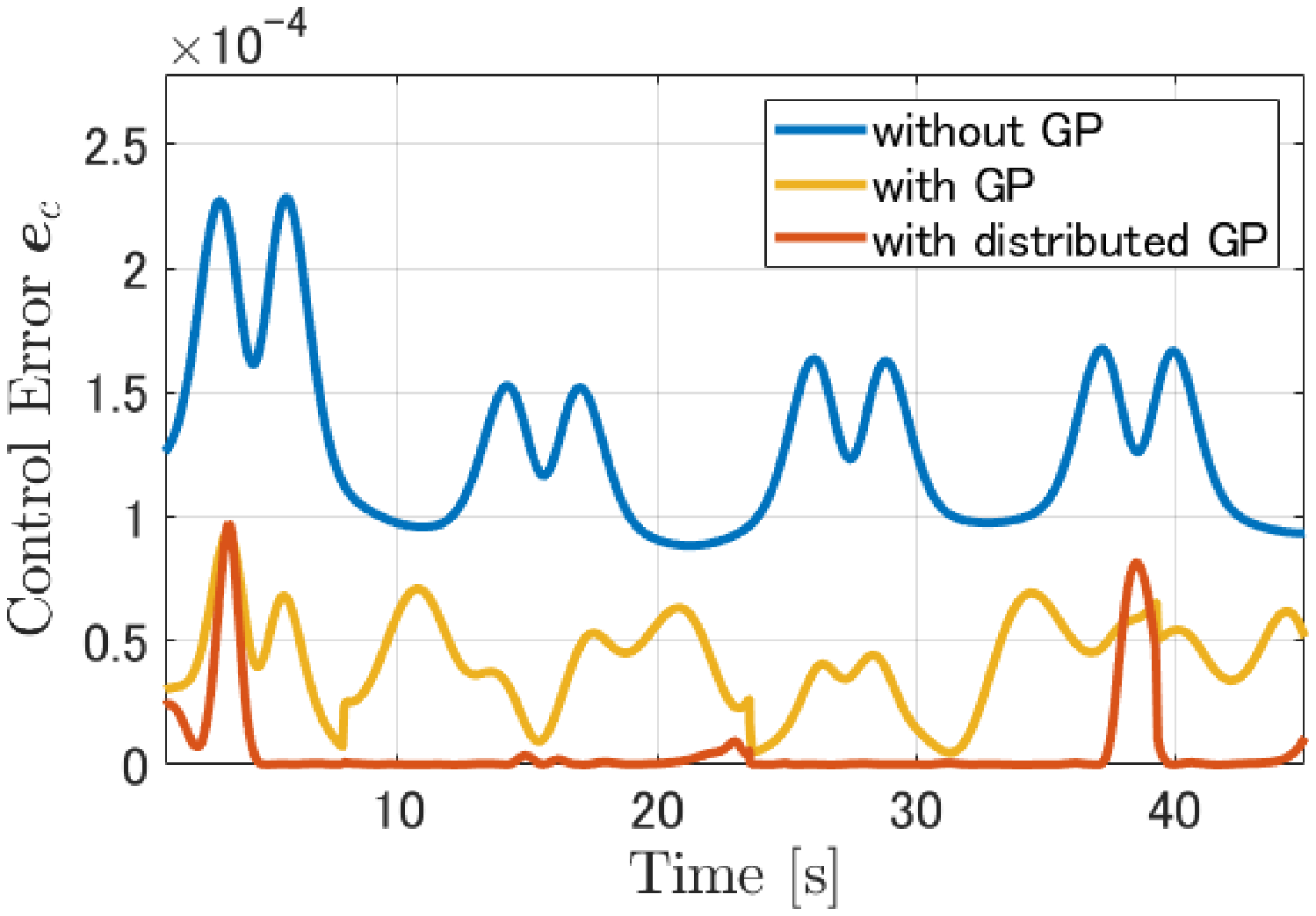}
    \end{center}
    \caption{Error in cooperative visual pursuit control with three drones}
    \label{fig:ec}
    \begin{center}
        \includegraphics[width=75mm]{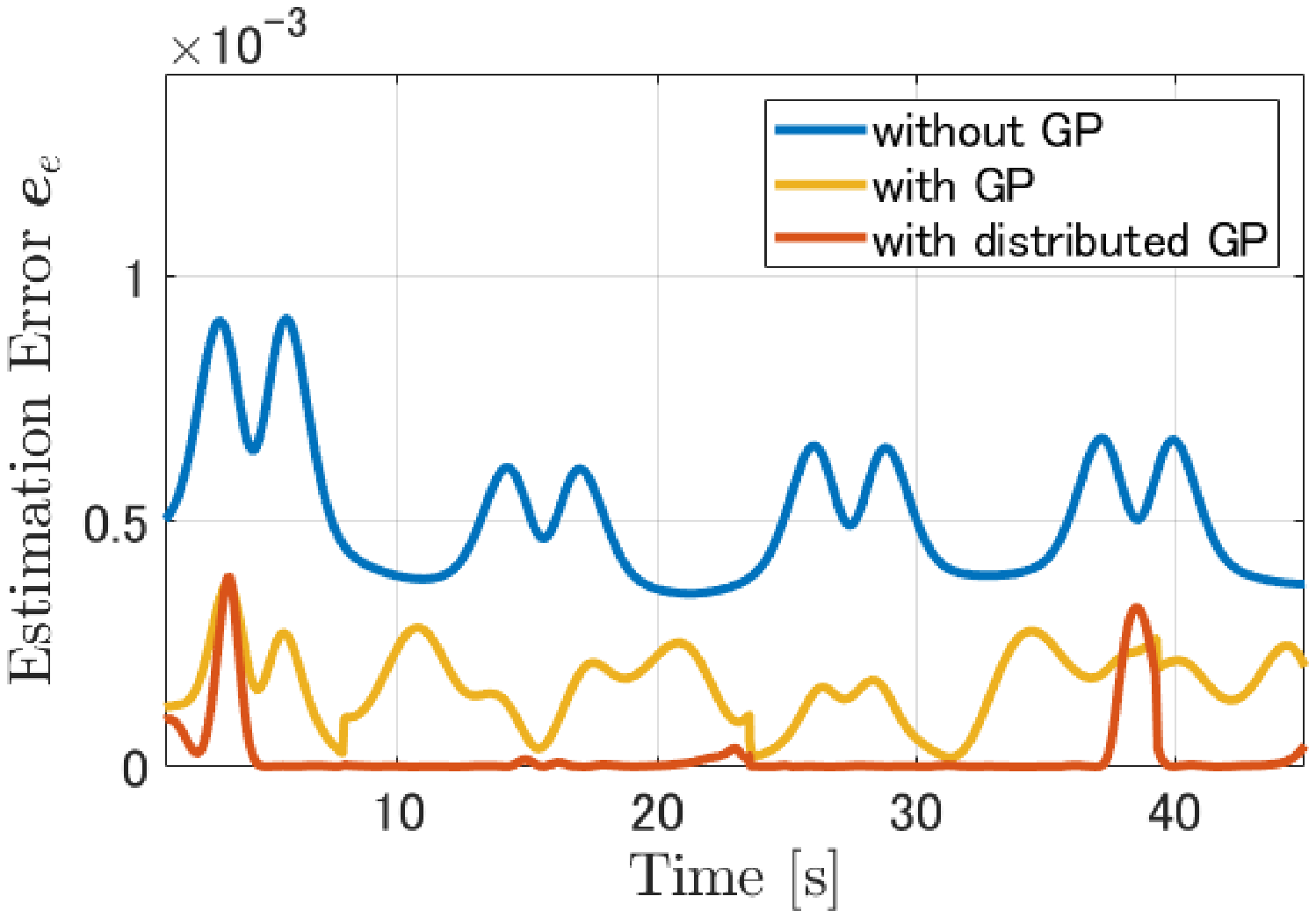}
    \end{center}
    \caption{Estimation error in cooperative visual pursuit control with three drones}
    \label{fig:ee}
\end{figure}

In this simulation, we performed three types of simulations: one in which drones do not possess Gaussian process models, one in which each drone uses only its own GP expert, and one in which distributed Gaussian process regression is used.
The results are shown in Figs. \ref{fig:ec}, \ref{fig:ee}.

The squared mean of $e$ in the simulation for three cases are obtained as $3.0 \times 10^{-7}, 3.6 \times 10^{-8}$, and $5.6 \times 10^{-9}$, respectively, and the errors are found to be the smallest when the distributed Gaussian process regression is used.

These results quantitatively confirm that control performance is improved when distributed Gaussian process regression is used.
This is thought to be due to the appropriate combination of predicted body velocity inferred from within the expert region of each drone on the drone network.

%% file: sections/section_experiment.tex
\section{experiment}
After the simulation, an actual experiment is conducted in which two Tello drones produced by DJI follow Scamper produced by REVAST. 
In this experiment, only one Drone can observe the target.

The object is controlled to move at $0.1$m/s on a square of $1.0$m per side centered at the origin.
The rotation velocity is also input so that the object makes one rotation in one round at this time.

In this experiment, the desired pose $g_d$ of $g_{wi}$ and the desired body velocity $v_d$ that should be input to the target existing at $g_d$ are calculated on the computer, and control is performed by inputting the body velocity as (\ref{equ:scamper_vel}).
\begin{equation}\label{equ:scamper_vel}
    \Vb{w0} = \Ad{-R(\theta_{w0})} (R(\theta_d)v_d - 0.05 vec(g_{w0} g_d^{-1}))
\end{equation}

For actual drone control, it is difficult to increase the gain due to the communication delay and the slow operating frequency of the drone.
Therefore, in this paper, each gain is set as follows: $k_{ci1} = k_{ci2} = k_{ci3} = 13.0,~ k_{ci4} = 7.0,~ k_{eij} = 8.0,~ k_s = 1.0$.
Feature point extraction is performed by attaching markers to the target and acquiring color information.

In the experiment, as in the simulation, the expert regions are divided and training data are acquired within each region, and prediction is performed by Gaussian process regression. 
The creation of training data and learning of hyperparameters are performed in the same way as in the simulation.

\begin{figure}
    \begin{center}
        \includegraphics[width=75mm]{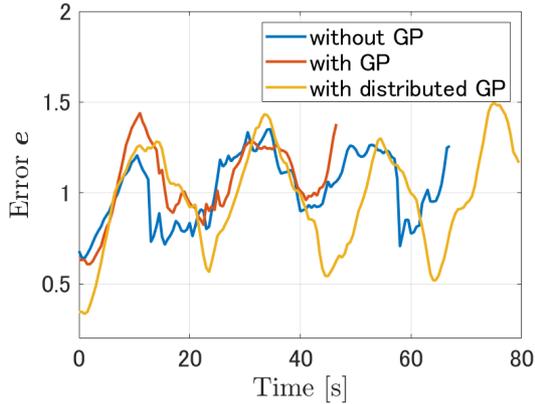}
    \end{center}
    \caption{Estimation and control error of position in three different experiments.}
    \label{fig:exp_result}
\end{figure}

\begin{figure}
    \begin{center}
        \includegraphics[width=0.75\linewidth]{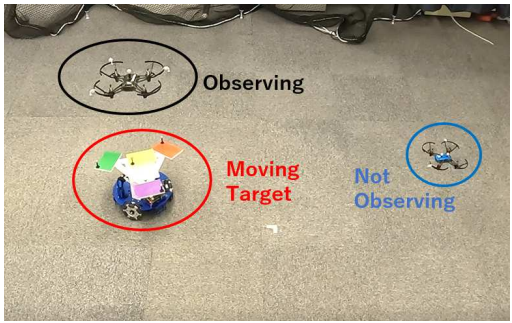}
    \end{center}
    \caption{Snapshot of the experiment.}
    \label{fig:exp}
\end{figure}

Under the above conditions, three types of experiments are conducted as well as simulations. 
The results are shown in Fig. \ref{fig:exp_result}.
Only control and estimation errors of position are plotted in the figure for simplicity.
The interrupted graph means that the target is no longer observable from the angle of view of the drone and that the target has been lost.
The experiment is shown in Fig. \ref{fig:exp}.
You can watch the video of the experiment with distributed Gaussian process from the link (\url{https://www.youtube.com/watch?v=a-3m45D50Vo}).

From the experiments, it is confirmed that the control method using distributed Gaussian process regression proposed in this paper works stably even in the real world.
This is supported by the fact that it did not lose the target, reduced the control and estimation errors more, and reduced oscillations more compared to the other two methods in the experiments.
This result may be because the dataset possessed by the neighboring drone are taken into account.
However, the gain used in the experiment does not satisfy the gain condition of the theorem \ref{thm:dGP}. 
In order to achieve the control that satisfies the theorem, it is necessary to improve the hardware such as the communication speed and the operating frequency of the drone, as well as to improve the image processing performance to reduce the delay and noise.

%% file: sections/section_conclusion.tex
\section{Conclusion}
In this paper, we propose a control law for camera-equipped drone networks to pursue a target rigid body with unknown motion.
The proposed method uses distributed Gaussian process regression instead of the conventional single Gaussian process regression to reduce the computational load on the drone, which is not equipped with a high performance computer, and thus enables real time implementation.
Numerical simulations and experiments show that when the drones possess the GP expert model, the control performance is better when the target body velocity is estimated by distributed Gaussian process regression than by Gaussian process regression by only each drone, indicating the effectiveness of the proposed method.